\documentclass[%
 reprint,
 amsmath,amssymb,
 aps,nofootinbib,
 prl,superscriptaddress
 ]{revtex4-2}

\usepackage{hyperref}

\usepackage[utf8]{inputenc}
\usepackage{comment}

\usepackage{graphicx}% Include figure files
\usepackage{dcolumn}% Align table columns on decimal point
\usepackage{bm}% bold math
\usepackage{multibib}

\newcommand{\ket}[1]{\ensuremath{| {#1} \rangle }}
\newcommand{\bra}[1]{\ensuremath{\langle {#1} |}}

\renewcommand{\vec}[1]{\bm{#1}}

\newcites{SM}{Supplementary Material Bibliography}

\begin{document}

\title{Inclusive hadronic decay rate of the $\tau$ lepton from lattice QCD:\\ 
the $\bar u s$ flavour channel and the Cabibbo angle}

\author{Constantia Alexandrou}
\affiliation{Department of Physics, University of Cyprus, 20537 Nicosia, Cyprus}
\affiliation{Computation-based Science and Technology Research Center, The Cyprus Institute, 20 Konstantinou Kavafi Street, 2121 Nicosia, Cyprus}

\author{Simone Bacchio}
%\email{}
\affiliation{Computation-based Science and Technology Research Center, The Cyprus Institute, 20 Konstantinou Kavafi Street, 2121 Nicosia, Cyprus}

\author{Alessandro De Santis}%
%\email{alessandro.desantis@roma2.infn.it}
\affiliation{Dipartimento di Fisica and INFN, Universit\`a di Roma Tor Vergata, Via della Ricerca Scientifica 1, I-00133 Roma, Italy}

\author{Antonio Evangelista}%
%\email{alessandro.desantis@roma2.infn.it}
\affiliation{Dipartimento di Fisica and INFN, Universit\`a di Roma Tor Vergata, Via della Ricerca Scientifica 1, I-00133 Roma, Italy}

\author{Jacob Finkenrath}
\affiliation{Bergische Universität Wuppertal, Gaußstraße 20, 42119 Wuppertal, Germany}

\author{Roberto Frezzotti}
%\email{}
\affiliation{Dipartimento di Fisica and INFN, Universit\`a di Roma Tor Vergata, Via della Ricerca Scientifica 1, I-00133 Roma, Italy}

\author{Giuseppe Gagliardi}
%\email{}
\affiliation{Istituto Nazionale di Fisica Nucleare, Sezione di Roma Tre, Via della Vasca Navale 84, I-00146 Rome, Italy}

\author{Marco Garofalo}
%\email{}
\affiliation{HISKP (Theory), Rheinische Friedrich-Wilhelms-Universit\"at Bonn,
	Nussallee 14-16, 53115 Bonn, Germany}

\author{Bartosz Kostrzewa}
%\email{}
\affiliation{High Performance Computing and Analytics Lab, Rheinische Friedrich-Wilhelms-Universit\"at Bonn, Friedrich-Hirzebruch-Allee 8, 53115 Bonn, Germany}

\author{Vittorio Lubicz}
%\email{}
\affiliation{Dipartimento di Matematica e Fisica, Universit\`a Roma Tre and INFN, Sezione di Roma Tre, Via della Vasca Navale 84, I-00146 Rome, Italy}

\author{Simone Romiti}
%\email{}
\affiliation{HISKP (Theory), Rheinische Friedrich-Wilhelms-Universit\"at Bonn,
	Nussallee 14-16, 53115 Bonn, Germany}

\author{Francesco Sanfilippo}
%\email{}
\affiliation{Istituto Nazionale di Fisica Nucleare, Sezione di Roma Tre, Via della Vasca Navale 84, I-00146 Rome, Italy}

\author{Silvano Simula}
%\email{}
\affiliation{Istituto Nazionale di Fisica Nucleare, Sezione di Roma Tre, Via della Vasca Navale 84, I-00146 Rome, Italy}

\author{Nazario Tantalo}
%\email{}
\affiliation{Dipartimento di Fisica and INFN, Universit\`a di Roma Tor Vergata, Via della Ricerca Scientifica 1, I-00133 Roma, Italy}

\author{Carsten Urbach}
\affiliation{
HISKP (Theory), Rheinische Friedrich-Wilhelms-Universit\"at Bonn,
	Nussallee 14-16, 53115 Bonn, Germany
}

\author{Urs Wenger}
\affiliation{Institute for Theoretical Physics, Albert Einstein Center for Fundamental Physics,\\University of Bern, Sidlerstrasse 5, CH-3012 Bern, Switzerland}

\collaboration{Extended Twisted Mass Collaboration (ETMC)}%\noaffiliation

\begin{abstract}
We present a lattice determination of the inclusive decay rate of the process $\tau\mapsto X_{us} \nu_\tau$ in which the $\tau$ lepton decays into a generic hadronic state $X_{us}$ with $\bar u s$ flavour quantum numbers. Our results have been obtained in $n_f=2+1+1$ iso-symmetric QCD with full non-perturbative accuracy, without any OPE approximation and, except for the presently missing long-distance isospin-breaking corrections, include a solid estimate of all sources of theoretical uncertainties. This has been possible by using the Hansen-Lupo-Tantalo method~\cite{Hansen:2019idp} that we have already successfully applied in~\cite{Evangelista:2023fmt} to compute the inclusive decay rate of the process $\tau\mapsto X_{ud} \nu_\tau$ in the $\bar u d$ flavour channel. By combining our first-principles theoretical results with the presently-available experimental data we extract the CKM matrix element $\vert V_{us}\vert$, the Cabibbo angle, with a $0.9$\% accuracy, dominated by the experimental error.  
\end{abstract}

\maketitle

%%%%%%%%%%%%%%%%%%%%%%%%%%%%%%%%%%%%%%%%%%%%%%%%%%%%%%%%%%%%%%%%%%%%%%%%%%%%%%%%%%%%%%%%%%%%%%%%
\section{
\label{sec:introduction}
Introduction
}
%%%%%%%%%%%%%%%%%%%%%%%%%%%%%%%%%%%%%%%%%%%%%%%%%%%%%%%%%%%%%%%%%%%%%%%%%%%%%%%%%%%%%%%%%%%%%%%%

The hadronic decays of the $\tau$ lepton represent very important probes of both the leptonic and hadronic flavour sectors of the Standard Model (SM). A particularly interesting test is the one associated with the Cabibbo angle, more precisely the CKM matrix element $\vert V_{us}\vert$, that can be extracted from both exclusive and inclusive hadronic $\tau$ decays and then compared with independent determinations coming from hadronic decays.
Currently, the most precise determinations of $\vert V_{us}\vert$ are obtained from semileptonic kaon decays, $\vert V_{us}\vert_{K_{\ell 3}}=0.2232(6)$, and from the ratio of the leptonic decay rates of kaons and pions, $\vert V_{us}\vert_{K/\pi_{\ell 2}} = 0.2254(5)$~\cite{FlavourLatticeAveragingGroupFLAG:2021npn, ParticleDataGroup:2022pth}. The two determinations exhibit a tension at the level of $2.8$ standard deviations (SD). 

The exclusive decay rate $\Gamma\left( \tau\mapsto K\nu_\tau \right)$ can be computed very precisely in QCD. Indeed,   
by neglecting long-distance QED radiative corrections, the non-perturbative input needed to compute $\Gamma\left( \tau\mapsto K\nu_\tau \right)$ is the same needed to compute the decay rate $\Gamma\left(K\mapsto \ell \bar\nu_\ell \right)$, namely the leptonic decay constant $f_K$. By combining the world-average of the lattice QCD results for $f_K$ given in Ref.~\cite{FlavourLatticeAveragingGroupFLAG:2021npn} with   the average of the presently available experimental measurements of $\Gamma\left( \tau\mapsto K\nu_\tau \right)$, Ref.~\cite{HFLAV:2022esi} quotes $\vert V_{us}\vert_{\tau\mathrm{-excl}}= 0.2219(17)$, a value that is well compatible ($0.7$ SD) with $\vert V_{us}\vert_{K_{\ell 3}}$, but lower ($2$ SD) than $\vert V_{us}\vert_{K/\pi_{\ell 2}}$.

The \emph{focus of this letter} are the inclusive decays of the $\tau$ in generic hadronic final states $X_{us}$ with $\bar u s$ flavour quantum numbers. We provide, for the first time, first-principles lattice results for the normalized decay rate
\begin{flalign}
R^{(\tau)}_{us}=
\frac{\Gamma(\tau\mapsto X_{us} \nu_\tau)}{\Gamma(\tau\mapsto e \bar\nu_e \nu_\tau)}\;,
\label{eq:Rusdefinition}
\end{flalign}
that we obtained in $n_f=2+1+1$ iso-symmetric QCD with full non-perturbative accuracy, without any Operator Product Expansion (OPE) approximation and that, except for the presently missing long-distance isospin-breaking corrections\footnote{The long-distance QED and strong isospin-breaking corrections (that are presently known only for a limited subset of the exclusive hadronic channels contributing to $\tau\mapsto X_{us} \nu_\tau$) have been neglected in all previous calculations of $R^{(\tau)}_{us}$.}, include a solid estimate of all sources of theoretical uncertainties.

$R^{(\tau)}_{us}$ is an inclusive quantity that depends upon an energy scale (the $\tau$ mass $m_\tau$) which is quite higher than $\Lambda_\mathrm{QCD}$ and has been extensively studied in the phenomenological literature by relying on asymptotic freedom and by using perturbative and/or OPE approximations. 
The OPE analysis performed in Ref.~\cite{Gamiz:2006xx,Pich:2013lsa}, and recently reviewed in Ref.~\cite{HFLAV:2022esi}, gives $\vert V_{us}\vert_{\tau\mathrm{-OPE-1}}= 0.2184(21)$. A different analysis, performed in Refs.~\cite{Hudspith:2017vew,Maltman:2019xeh} by determining the higher order terms in the OPE expansion by fits to lattice current-current correlators and by using a partly different experimental input, gives $\vert V_{us}\vert_{\tau\mathrm{-OPE-2}}= 0.2219(22)$. While these two results are compatible at the level of $1$ SD, in fact, $\vert V_{us}\vert_{\tau\mathrm{-OPE-1}}$ is in strong tension ($3.2$ SD) with $\vert V_{us}\vert_{K/\pi_{\ell 2}}$, see FIG.~\ref{fig:comp_Vus}.

A direct non-perturbative lattice calculation of $R^{(\tau)}_{us}$ has been deemed impossible for several years because of the problem associated with the extraction of the needed non-perturbative physical input, i.e.\ the spectral density of two hadronic weak currents, from the corresponding lattice current-current correlators. 

The problem has been circumvented in Ref.~\cite{RBC:2018uyk} by targeting the calculation of spectral integrals that can readily be obtained starting from the lattice current-current correlators. While the method avoids OPE assumptions, it requires perturbative inputs to extract $\vert V_{us}\vert$ from the dispersion integral of the measured hadronic $\tau$ decays that, in fact, do not provide experimental information on the hadronic spectral density for energies larger than $m_\tau$. The considered dispersion relations have been tailored to minimize the impact of these perturbative inputs and (taking into account the update of Refs.~\cite{Maltman:2019xeh}) the result $\vert V_{us}\vert_{\tau\mathrm{-latt-disp}}=0.2240(18)$ has been obtained.
The nice agreement of $\vert V_{us}\vert_{\tau\mathrm{-latt-disp}}$ with both $\vert V_{us}\vert_{K/\pi_{\ell 2}}$ and $\vert V_{us}\vert_{\tau\mathrm{-excl}}$ can be traced back to the fact, emphasized in Ref.~\cite{Crivellin:2022rhw}, that the particular dispersion relation used to get $\vert V_{us}\vert_{\tau\mathrm{-latt-disp}}$ mostly relies on the exclusive decay $\tau\mapsto K \nu_\tau$ (which instead contributes for less than $25$\% to $R^{(\tau)}_{us}$).

In Ref.~\cite{Evangelista:2023fmt}, by building on previous ideas~\cite{Hansen:2017mnd,Gambino:2020crt}, we have shown that a direct non-perturbative lattice  calculation of inclusive hadronic decay rates of the $\tau$ is possible by using the HLT method of Ref.~\cite{Hansen:2019idp} for the extraction of smeared spectral densities from lattice correlators. 
In that companion paper we provided all the theoretical ingredients needed to directly extract $R^{(\tau)}_{us}$ from the current-current lattice correlators and performed the first non-perturbative calculation of $R^{(\tau)}_{ud}$, i.e.\ the normalized inclusive decay rate in the $\bar u d$ flavour channel. By combining our first principles lattice result $R^{(\tau)}_{ud}/\vert V_{ud}\vert^2=3.650(28)$ with the world average of the experimental data given in Ref.~\cite{HFLAV:2022esi} we obtained $\vert V_{ud}\vert_{\tau\mathrm{-latt-incl}}=0.9752(39)$. Our result for $\vert V_{ud}\vert_{\tau\mathrm{-latt-incl}}$ has a $0.4$\% error and is fully compatible with the more precise result $\vert V_{ud}\vert_{0^{+}}= 0.97373 (31)$ coming from superallowed nuclear $\beta$-decays~\cite{Hardy:2020qwl}. 

In this work we apply the method of Ref.~\cite{Evangelista:2023fmt} in the $\bar u s$ flavour channel and present our first-principles lattice QCD result 
\begin{flalign}
R^{(\tau)}_{us}/\vert V_{us}\vert^2=3.407(22)\;.
\end{flalign}
From this, by using the world-average of the experimental data given in Ref.~\cite{HFLAV:2022esi}, we get
\begin{flalign}
\vert V_{us}\vert_{\tau\mathrm{-latt-incl}}=0.2189(7)_\mathrm{th}(18)_\mathrm{exp}\;.
\end{flalign}
Our result, being in very good agreement with both $\vert V_{us}\vert_{\tau\mathrm{-OPE-1}}$ and $\vert V_{us}\vert_{\tau\mathrm{-OPE-2}}$, confirms the previous estimates of $\vert V_{us}\vert$ from inclusive hadronic $\tau$ decays and, therefore, also confirms the previous observed tension of about $2$\,-\,$3~$SD w.r.t. other determinations.

%%%%%%%%%%%%%%%%%%%%%%%%%%%%%%%%%%%%%%%%%%%%%%%%%%%%%%%%%%%%%%%%%%%%%%%%%%%%%%%%%%%%%%%%%%%%%%%%
\section{
\label{sec:materialsandmethods}
Methods and Materials 
}
%%%%%%%%%%%%%%%%%%%%%%%%%%%%%%%%%%%%%%%%%%%%%%%%%%%%%%%%%%%%%%%%%%%%%%%%%%%%%%%%%%%%%%%%%%%%%%%%

\paragraph{Methods.}The method for a direct lattice QCD calculation of $R^{(\tau)}_{us}/\vert V_{us}\vert^2$ has been introduced and explained in full details in Ref.~\cite{Evangelista:2023fmt}. 
The starting point of the calculation is the following representation of the normalized inclusive decay rate
\begin{flalign}
&
\frac{m_\tau^3\, R^{(\tau)}_{us}}{12\pi S_\mathrm{EW}\vert V_{us}\vert^2}
=
\lim_{\sigma\mapsto 0}
\sum_{\mathrm{I}=\mathrm{T},\mathrm{L}}
\int_0^\infty dE
K_\mathrm{I}^\sigma\left(\frac{E}{m_\tau}\right)E^2\rho_\mathrm{I}(E^2)
\label{eq:representaion1}
\end{flalign}
which we are now going to illustrate.
The factor $S_\mathrm{EW}=1.0201(3)$ takes into account the short-distance electroweak corrections~\cite{Erler:2002mv}. The scalar form factors $\rho_\mathrm{T}$ and $\rho_\mathrm{L}$ are the transverse ($\mathrm{T}$) and longitudinal ($\mathrm{L}$) components of the hadronic spectral density
\begin{flalign}
\rho_{us}^{\alpha\beta}(q)
&=
\bra{0} J_{us}^\alpha(0)\, (2\pi)^4 \delta^4(\mathbb{P}-q)\, J_{us}^\beta(0)^\dagger \ket{0}
\nonumber \\[4pt]
&= \left(q^\alpha q^\beta-g^{\alpha\beta}q^2 \right)\rho_\mathrm{T}(q^2)
+
q^\alpha q^\beta \rho_\mathrm{L}(q^2)
\;,
\end{flalign}
where $\mathbb{P}$ is the QCD 4-momentum operator and $J_{us}^\alpha=\bar u \gamma^\alpha(1-\gamma^5) s$ is the hadronic weak current. These also appear in the spectral representation of the following current-current Euclidean correlator
\begin{flalign}
\label{eq:curr_curr_corr}
C^{\alpha\beta}(t)
=\int d^3x\, T\bra{0} J_{us}^\alpha(t,\vec x)\, J_{us}^\beta(0)^\dagger\ket{0}\;.
\end{flalign}
Indeed,
\begin{flalign}
&
C_\mathrm{T}(t)
=\frac{1}{3}\sum_{i=1}^3C^{ii}(t)
=
\int_0^\infty \frac{dE}{2\pi} e^{-Et}\, E^2\rho_\mathrm{T}(E^2)\;,
\nonumber \\[4pt]
&
C_\mathrm{L}(t)
=C^{00}(t)
=
\int_0^\infty \frac{dE}{2\pi} e^{-Et}\, E^2\rho_\mathrm{L}(E^2)\;.
\end{flalign}
The kernels 
\begin{flalign}
&
K_\mathrm{L}^\sigma(x)
=\frac{(1-x^2)^2\, \Theta_\sigma(1-x)}{x}\;,
\nonumber \\[4pt]
&
K_\mathrm{T}^\sigma(x)
=(1+2x^2)\, K_\mathrm{L}^\sigma(x)\;,
\nonumber \\[4pt]
&
\Theta_\sigma(x)=\frac{1}{1+e^{-\frac{x}{\sigma}}}\;,
\qquad
\lim_{\sigma\mapsto 0}\Theta_\sigma(x) = \theta(x)\;,
\end{flalign}
are proportional to the phase-space factors and to an infinitely differentiable smooth representation of the Heaviside step function $\theta(x)$ introduced in order to be able to apply the HLT method~\cite{Hansen:2019idp}. 
In the limit in which the smearing parameter $\sigma$ vanishes the energy integral of Eq.~(\ref{eq:representaion1}) is restricted to the physical range $E\in [0,m_\tau]$. 

As explained in full details in Ref.~\cite{Evangelista:2023fmt}, it is possible to obtain at finite lattice spacing ($a$) approximate representations of the kernels $K_\mathrm{I}^\sigma(E/m_\tau)$, 
\begin{flalign}
\tilde K_\mathrm{I}^\sigma\left(\frac{E}{m_\tau};\vec g_\mathrm{I}\right)
=\sum_{n=1}^{N} g_\mathrm{I}(n) e^{-naE} \;,
\end{flalign}
in terms of the coefficients $\vec g_\mathrm{I}$. The error of this approximation can be made to vanish in the limit of an infinite number of euclidean lattice times ($N\mapsto \infty$). The HLT method provides at finite $N$ coefficients $\vec g_\mathrm{I}^\star$ corresponding to optimal representations 
\begin{flalign}
\frac{m_\tau^3\, R^{(\tau,\mathrm{I})}_{us}(\sigma)}{24\pi^2 S_\mathrm{EW}\vert V_{us}\vert^2}
= 
\sum_{n=1}^{N} g_\mathrm{I}^\star(n)\, C_\mathrm{I}(an)
\label{eq:sumresult}
\end{flalign}
of the smeared spectral integrals appearing in Eq.~(\ref{eq:representaion1}) so that, up to statistical and systematic errors,
\begin{flalign}
R^{(\tau)}_{us}
=\lim_{\sigma\mapsto 0}
\sum_{\mathrm{I}=\mathrm{T},\mathrm{L}}
R^{(\tau,\mathrm{I})}_{us}(\sigma)
\;.
\end{flalign}
See the Appendix and  Ref.~\cite{Evangelista:2023fmt} for further details.

\paragraph{Materials.}The lattice gauge ensembles used in this work, generated by the Extended Twisted Mass Collaboration (ETMC), are listed in TABLE~\ref{tab:ensembles} and described in full details in Ref.~\cite{ExtendedTwistedMass:2022jpw}. With respect to that analysis we have included two additional gauge ensembles, the C112 and the E112 (the ensemble with the finest lattice spacing among those so-far produced by the ETMC). Moreover, we have computed the small corrections in the lattice bare parameters required to match the iso-symmetric QCD world defined by $f_{\pi} = 130.5$~MeV, $m_{\pi} = 135.0$~MeV, $m_{K} = 494.6$~MeV and $m_{D_{s}} = 1967$~MeV. This explains the small difference between the lattice spacings and renormalization constants given in TABLE~\ref{tab:ensembles} and the ones quoted in Ref.~\cite{ExtendedTwistedMass:2022jpw}.

We relied on the same mixed-action setup described in Refs.~\cite{ExtendedTwistedMass:2022jpw,Frezzotti:2004wz} and evaluated, for each of the ensembles in TABLE~\ref{tab:ensembles}, the current-current correlator in Eq.~(\ref{eq:curr_curr_corr}), extending to the $\bar{u}s$ flavour channel the calculation performed in Ref.~\cite{Evangelista:2023fmt} in the $\bar{u}d$ sector (to which we refer for further technical details). In full analogy with that calculation, we considered two different regularizations of the weak hadronic current $J_{us}^{\alpha}$, which give rise to the the so-called Twisted Mass (``tm'') and Osterwalder-Seiler (``OS'') lattice correlators $C^{\alpha\beta}(t)$. The results for $R_{us}^{(\tau)}(\sigma)$ obtained in the two regularizations differ by $O(a^2)$   cutoff effects~\cite{Frezzotti:2003ni,Frezzotti:2005gi} and must coincide in the continuum limit.

\begin{table}[t]
\begin{ruledtabular}
\begin{tabular}{lccccc}
\textrm{ID}&
$L/a$&
$a$ \textrm{fm}&
$L$ \textrm{fm}&
$Z_{V}$ & $Z_{A}$ \\
\colrule
\textrm{B64} & $64$ & 0.07951(4) & 5.09 & 0.706377(20) & 0.74300(21) \\
\textrm{B96} & $96$ & 0.07951(4) & 7.63 & 0.706427(10) & 0.74278(20)  \\
\textrm{C80} & $80$ & 0.06816(8) & 5.45 & 0.725405(14) & 0.75814(13) \\
\textrm{C112} & $112$ & 0.06816(8) & 7.63 & 0.725421(10) & 0.75828(11)  \\
\textrm{D96} & $96$ & 0.05688(6) & 5.46 & 0.744110(7) & 0.77367(8)  \\
\textrm{E112} & $112$ & 0.04891(6) & 5.48 & 0.758231(5) &  0.78542(7)
\end{tabular}
\end{ruledtabular}
\caption{\label{tab:ensembles}
ETMC gauge ensembles used in this work.
We give the values of the lattice spacing $a$, of the spatial lattice extent $L$, and of the vector and axial renormalization constants $Z_{V}$ and $Z_{A}$. The temporal extent of the lattice is always $T=2L$.}
\end{table}

%%%%%%%%%%%%%%%%%%%%%%%%%%%%%%%%%%%%%%%%%%%%%%%%%%%%%%%%%%%%%%%%%%%%%%%%%%%%%%%%%%%%%%%%%%%%%%%%
\section{
\label{sec:results}
Results
}
%%%%%%%%%%%%%%%%%%%%%%%%%%%%%%%%%%%%%%%%%%%%%%%%%%%%%%%%%%%%%%%%%%%%%%%%%%%%%%%%%%%%%%%%%%%%%%%%
%

In our calculation we considered several values of the smearing parameter $\sigma \in [0.01,0.16]$, and evaluated $R_{us}^{(\tau,I)}(\sigma)$ on all the ensembles of TABLE~\ref{tab:ensembles} by using the HLT method. The results of the HLT analyses at $\sigma>0$, including a quantitative study of the finite-size effects, are presented and discussed in the Appendix. Here below we discuss the continuum and $\sigma\mapsto 0$ extrapolations from which we obtain our physical result for $R_{us}^{(\tau)}$. Further technical details on the analysis procedure can be found in Ref.~\cite{Evangelista:2023fmt}.

\begin{figure}
\includegraphics[scale=0.22]{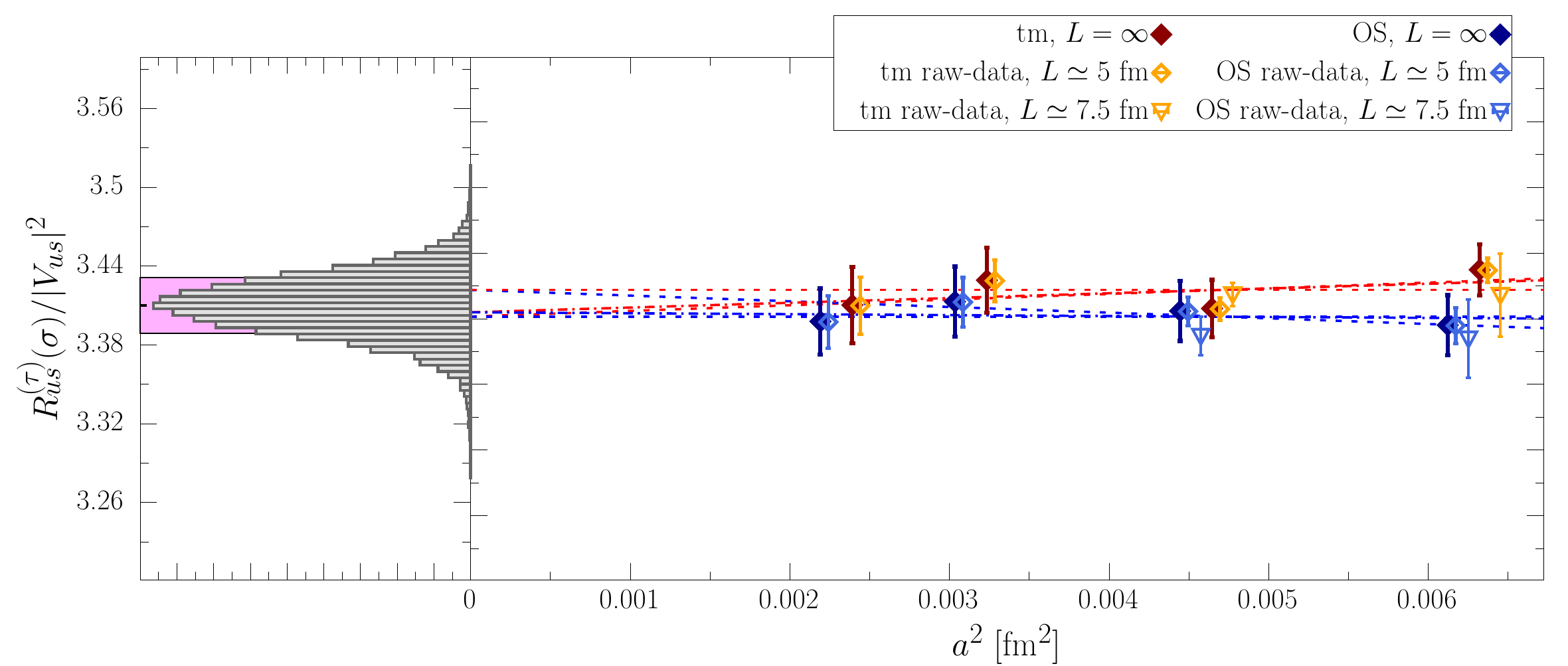}
\caption{\label{fig:continuum_extr} Illustrative example of the continuum extrapolation of $R_{us}^{(\tau)}(\sigma)$ for $\sigma=0.02$. The data points in light-blue and orange correspond to the raw data obtained on the ensembles listed in TABLE~\ref{tab:ensembles} respectively for the ``OS'' and ``tm'' regularizations. The data points in dark-red and dark-blue are instead inclusive of the systematic error due to finite-size effects. The different red (for `tm'')  and blue (for `OS'') lines show some of the fits obtained using a constant or linear Ansatz in $a^{2}$. The histogram shown in the left part of the figure corresponds to the distribution of the continuum extrapolated results obtained after applying the BAIC. All data correspond to the kernel reconstructions obtained with the choice $\alpha=r_{max}=4$ of the HLT algorithmic parameters (see Appendix).}
\end{figure}

In FIG.~\ref{fig:continuum_extr}, we give an example of the continuum extrapolation for $R_{ud}^{(\tau)}(\sigma)$, which we perform separately for each simulated value of $\sigma$. 
To perform the extrapolations, we take advantage of the fact that in the continuum limit the results corresponding to the ``tm'' and ``OS'' regularizations must coincide, and thus perform a combined extrapolation of the form
\begin{align}
R_{us}^{(\tau)}(\sigma, {\rm tm}) &= R(\sigma) + D_1^{{\rm tm}}(\sigma) a^{2} + D_{2}^{{\rm tm}}(\sigma) a^{4}~,\\
R_{us}^{\tau}(\sigma, {\rm OS}) &= R(\sigma) + D_1^{\rm OS}(\sigma) a^{2} + D_{2}^{\rm OS}(\sigma) a^{4}~,
\end{align}
where $R(\sigma)$, $D_1^{\rm tm/OS}(\sigma)$, and $D_{2}^{\rm tm/OS}(\sigma)$ are $\sigma-$dependent free fit parameters. We perform both constant, linear and quadratic extrapolations in $a^{2}$. At small values of $\sigma \lesssim 0.12$, where the size of the cut-off effects is remarkably small, we did not perform fits including the $a^{4}$ terms. In order to combine the results obtained in the different correlated continuum fits, and provide our final determination of $R_{us}^{(\tau)}(\sigma)$, we make use of the Bayesian Akaike Information Criterion (BAIC) discussed in section III.B of Ref.~\cite{Evangelista:2023fmt}. The histogram shown in FIG.~\ref{fig:continuum_extr} corresponds to the p.d.f. of the continuum extrapolated results. For all $\sigma$ we checked that at least one of the fits performed has a $\chi^{2}/dof$ close to unit. To provide a quantitative measure of the quality of our continuum-limit extrapolations, we considered the spread
\begin{align}
\Delta_{a}(\sigma)=\frac{\left\vert R_{us}^{(\tau)}(\sigma)-R_{us}^{(\tau)}(\sigma, a^\mathrm{min})\right\vert
}{\Delta R_{us}^{(\tau)}(\sigma)}
\end{align}
between the continuum extrapolated value of $R_{us}^{(\tau)}(\sigma)$ and the corresponding value at the finest simulated lattice spacing (ensemble E112), in units of the uncertainty of the continuum extrapolation $\Delta R_{us}^{(\tau)}(\sigma)$. The lattice spacing dependence is essentially absent within uncertainties for $\sigma < 0.1$, where we have $\Delta_{a}(\sigma)<0.1$, while it becomes increasingly pronounced by increasing $\sigma$.

\begin{figure}
\includegraphics[scale=0.28]{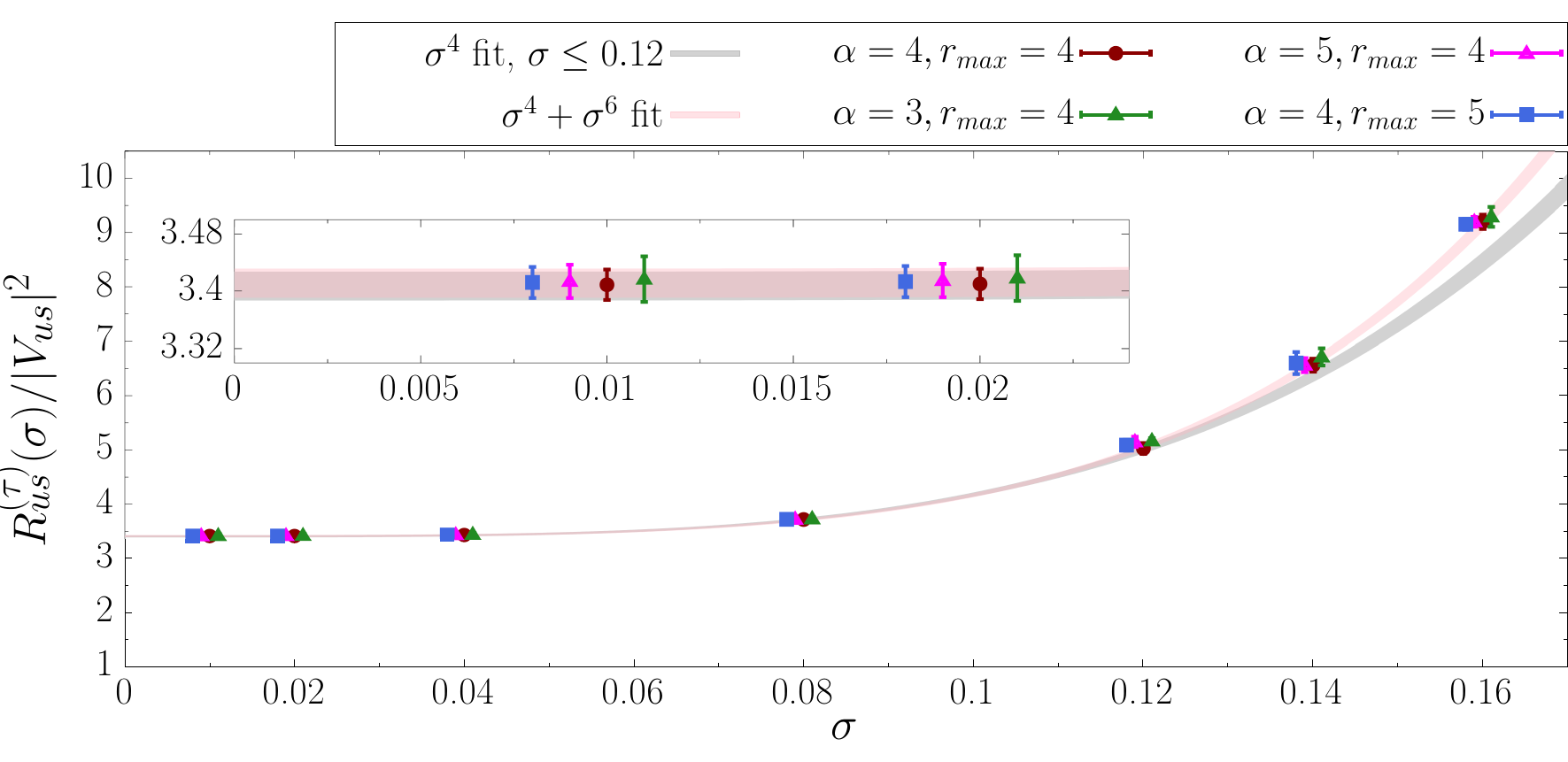}
\caption{\label{fig:sigma_extr} Extrapolation to vanishing $\sigma$. The gray and pink bands correspond to the $\sigma^{4}$ and $\sigma^{4}+\sigma^{6}$ fits to the data obtained by using $\alpha=r_{max}=4$ for the HLT algorithmic parameters (see Appendix). In the case of the $\sigma^{4}$ fit the data points at $\sigma > 0.12$ have been excluded. The results corresponding to different choices of the HLT algorithmic parameters are in remarkable good agreement.}
\end{figure}

To obtain our final determination of $R_{us}^{(\tau)}/|V_{us}|^{2}$, we need to perform the extrapolation to vanishing $\sigma$. According to the theoretical analysis presented in appendix~B of Ref.~\cite{Evangelista:2023fmt}, the corrections to the $\sigma=0$ limit are of the form
\begin{align}
R_{us}^{(\tau)}(\sigma) = R_{us}^{(\tau)} + R_{4}\,\sigma^{4} + \mathcal{O}(\sigma^{6})~.
\end{align}
To carry out the extrapolation and to properly estimate the associated systematic error, we perform a first fit to our data including only $\sigma^{4}$ corrections and considering all values of $\sigma \leq 0.12$, and a second, additional, $\sigma^{4}+\sigma^{6}$ fit over the full range of $\sigma$ explored. The results of these extrapolations are shown in FIG.~\ref{fig:sigma_extr}.
The $\mathcal{O}(\sigma^{6})$ corrections become numerically subleading for $\sigma \leq 0.12$, while the $\sigma^{4}$ corrections are subleading for $\sigma \leq 0.04$, where the quality of our continuum extrapolations are remarkably good and the dependence upon $\sigma$ is basically absent. 
Such behaviour allows us to take the $\sigma \mapsto 0$ limit with full confidence. 

FIG.~\ref{fig:sigma_extr} also shows that the results corresponding to different choices of the HLT algorithmic parameters (see Appendix) are in perfect agreement, thus confirming the reliability of our estimates of the systematic errors associated with the HLT reconstruction of the smearing kernels. 

Taking into account all sources of uncertainties, our final determination of $R_{us}^{(\tau)}/|V_{us}|^{2}$ is
\begin{align}
R_{us}^{(\tau)}/|V_{us}|^{2} &= 3.407 \, (19)_{\rm stat+HLT+FSE}(10)_{a}(4)_{\rm \sigma}
\nonumber \\ 
&= 3.407\,(22)~.
\end{align}
The first source of uncertainty is due to statistical errors, FSEs and also includes the systematic uncertainties associated with the HLT spectral resonstructions\footnote{The HLT and FSE systematic errors have been estimated with a data-driven approach (see Eq.~(\ref{eq:syst_HLT})) and therefore are entangled with the statistical error. Approximately, the HLT systematic error is negligible with respect to the stat and FSE contributions which are instead of similar size.}. The second source of uncertainty is due to the continuum-limit extrapolation and has been estimated by taking into account the spread between the results obtained in the different fits using the BAIC (see Eqs.~(46)-(47) of Ref.~\cite{Evangelista:2023fmt} for details). The third source of uncertainty is due to the $\sigma\mapsto 0$ extrapolation and it is given by the difference between the results obtained in the $\sigma^{4}$ and $\sigma^{4}+\sigma^{6}$ fits shown in FIG.~\ref{fig:sigma_extr}. By combining our theoretical result with the experimental result $R_{us}^{(\tau)}= 0.1632(27)$ quoted in Ref.~\cite{HFLAV:2022esi} we obtain
\begin{flalign}
\label{eq:Vus}
\vert V_{us}\vert_{\tau\mathrm{-latt-incl}}=0.2189(7)_\mathrm{th}(18)_\mathrm{exp}\;.
\end{flalign}

\begin{figure}
\includegraphics[scale=0.33]{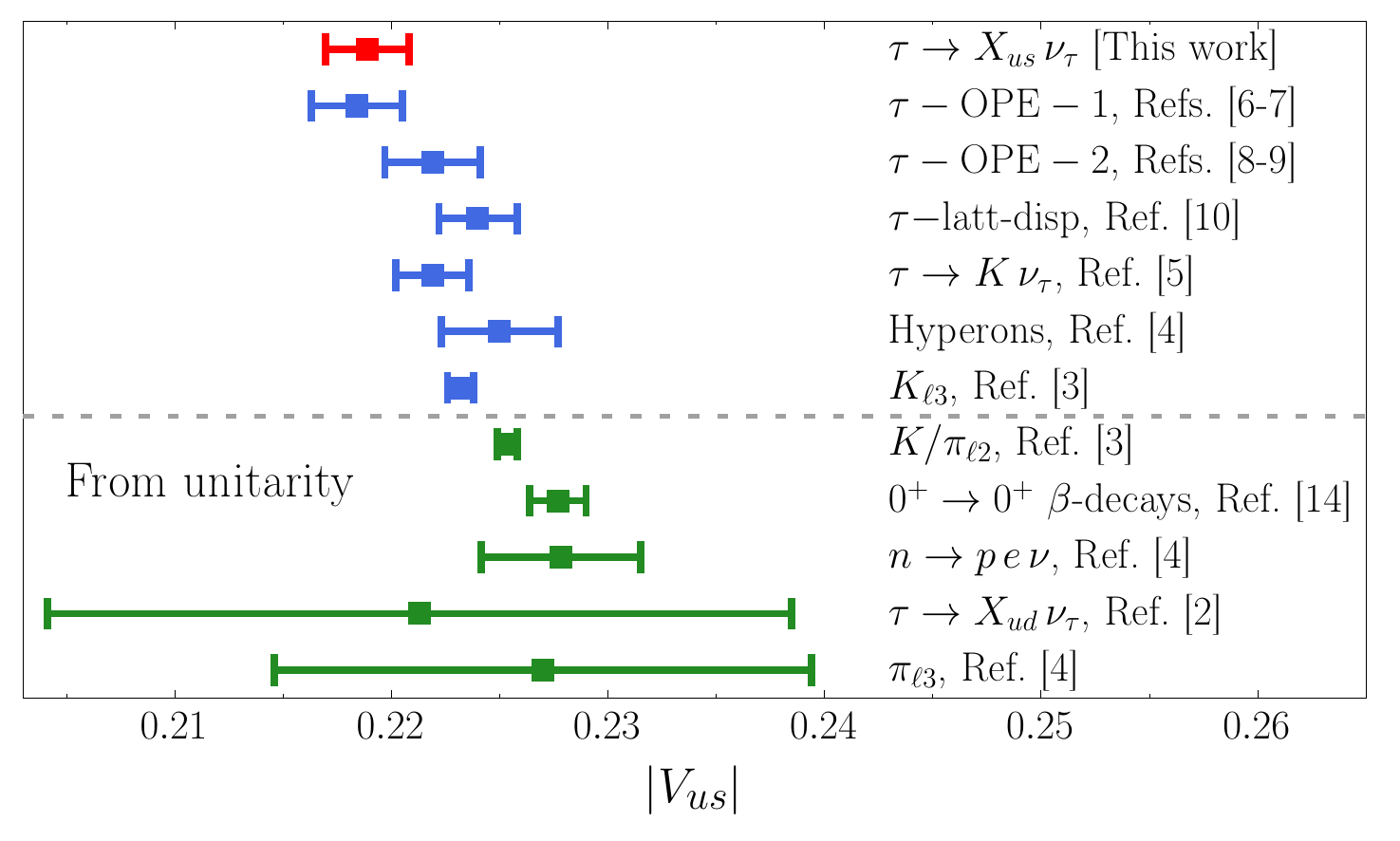}
\caption{\label{fig:comp_Vus} Comparison between our determination of $|V_{us}|$ (red data-point) and existing estimates based on $\tau$-decay analyses, or from other decay channels. The lower part of the figure shows the predictions for $|V_{us}|$ obtained assuming CKM-unitarity.}
\end{figure}

In FIG.~\ref{fig:comp_Vus} we compare our determination of $|V_{us}|$ 
with the other existing direct determinations as well as with various determinations obtained by assuming the unitarity of the CKM matrix, i.e. $|V_{us}| = \sqrt{ 1 - |V_{ud}|^{2}}$.
As the figure shows, our determination of $|V_{us}|$ from inclusive $\tau$ decay is in good agreement with both $|V_{us}|_{\tau-{\rm OPE}-1}$ and $|V_{us}|_{\tau-{\rm OPE}-2}$, while it is smaller (of about $2$ SD) than the determination of Ref.~\cite{RBC:2018uyk} which, however, mostly relies on the experimental value of the exclusive $\tau \to K \nu_{\ell}$ decay. 

Our current estimate of $\vert V_{us} \vert$ has been obtained by neglecting long distance isospin breaking corrections. These, instead, have been taken into account in the determinations $\vert V_{us}\vert_{K/\pi_{\ell 2}}$ and $\vert V_{us} \vert_{K_{\ell 3}}$  from %kaons and pions
leptonic and semileptonic decays~\cite{Cirigliano:2001mk,Cirigliano:2008wn,Cirigliano:2011tm,Giusti:2017dwk,DiCarlo:2019thl,Seng:2021boy,Seng:2021wcf,Seng:2022wcw,Boyle:2022lsi, Cirigliano:2022yyo}. The current difference between our result in Eq.~(\ref{eq:Vus}) and the determinations of $\vert V_{us} \vert$ from leptonic and semileptonic decays is at the level of $3.3$ and $2.2
$ SD, respectively. We note that in order to fully reconcile the $3.3$ SD difference w.r.t.\ $\vert V_{us} \vert_{K/\pi_{\ell 2}}$ one needs an isospin breaking correction 
\begin{flalign}
\delta R_{us}^{(\tau)}=
2\left\{
\frac{\vert V_{us}\vert_{\tau\mathrm{-latt-incl}}}{\vert V_{us}\vert_{K/\pi_{\ell 2}}}-1\right\}= -0.058(18)
\end{flalign}
on $R_{us}^{(\tau)}$.  
At the current level of the theoretical precision a first principles calculation of $\delta R_{us}^{(\tau)}$ on the lattice is needed. Once this calculation will be performed, experimental uncertainties will wholly govern the determination of $\vert V_{us} \vert$ from inclusive $\tau$ decays.
%%%%%%%%%%%%%%%%%%%%%%%%%%%%%%%%%%%%%%%%%%%%%%%%%%%%%%%%%%%%%%%%%%%%%%%%%%%%%%%%%%%%%%%%%%%%%%%%
\section{
\label{sec:conclusions}
Conclusions
}
%%%%%%%%%%%%%%%%%%%%%%%%%%%%%%%%%%%%%%%%%%%%%%%%%%%%%%%%%%%%%%%%%%%%%%%%%%%%%%%%%%%%%%%%%%%%%%%%
In this work we have extracted for the first time $\vert V_{us} \vert$ from inclusive hadronic $\tau$ decays with full non-perturbative accuracy and with a 0.9\% relative error that, currently, is dominated by the experimental uncertainty. 

Our iso-symmetric QCD result has been obtained without any perturbative approximation but is in fairly good agreement with previous estimates obtained by using OPE techniques. 
%, while it is smaller than the lattice result of Ref.~\cite{RBC:2018uyk} of about $2~$SD. 
Therefore, our result confirms the previously observed tension of about 3~SD between $\tau$-inclusive and purely hadronic determinations of $\vert V_{us} \vert$ which can no longer be attributed to the OPE approximation. 

The origin of this tension can possibly be ascribed to the long distance isospin breaking corrections, that have been taken into account in the determinations of $\vert V_{us} \vert$ coming from kaons and pions leptonic decays
but that, as in all previous determinations coming  from inclusive hadronic $\tau$ decays, we have presently neglected.  In fact, having obtained a fully non-perturbative result with sub-percent accuracy in iso-summetric QCD, further progress on the study of inclusive hadronic $\tau$ decays can only be done by computing these corrections from first principles. We have already started a series of projects dedicated to this challenging task.

On the other hand, we also noticed that in order to fully reabsorb the observed tension a rather large (of the order of 5\%) isospin breaking correction would be needed. In the light of this observation we think that it is important to investigate the possibility that experimental uncertainties on the $\tau$ inclusive hadronic decay rate have been underestimated and, at the same time, to speculate about possible new physics scenarios that could explain this puzzle.  

%%%%%%%%%%%%%%%%%%%%%%%%%%%%%%%%%%%%%%%%%%%%%%%%%%%%%%%%%%%%%%%%%%%%%%%%%%%%%%%%%%%%%%%%%%%%%%%%
\subsection{Acknowledgments}
\begin{acknowledgments}

\noindent This work, a small contribution to the long and exciting history of the mixing of quarks that he started 60 years ago, is dedicated to the memory of N.~Cabibbo.

The authors gratefully acknowledge the Gauss Centre for Supercomputing e.V. (www.gauss-centre.eu) for funding this project by providing computing time on the GCS Supercomputers SuperMUC-NG at Leibniz Supercomputing Centre and JUWELS~\cite{JUWELS} at Juelich Supercomputing Centre.

The authors acknowledge the Texas Advanced Computing Center (TACC) at The University of Texas at Austin for providing HPC resources (Project ID PHY21001).

The authors gratefully acknowledge PRACE for awarding access to HAWK at HLRS within the project with Id Acid 4886.

We gratefully acknowledge the Swiss National Supercomputing Centre (CSCS) and the EuroHPC Joint Undertaking for awarding this project access to the LUMI supercomputer, owned by the EuroHPC Joint Undertaking, hosted by CSC (Finland) and the LUMI consortium through the Chronos programme under project IDs CH17-CSCS-CYP and CH21-CSCS-UNIBE as well as the EuroHPC Regular Access Mode under project ID EHPC-REG-2021R0095.

%%%%%%%%%%%%%%%%%%%%%%%%%%
We gratefully acknowledge CINECA and EuroHPC JU for awarding this project access to Leonardo supercomputing hosted at CINECA.
We gratefully acknowledge CINECA for the provision of GPU time under the specific initiative INFN-LQCD123 and IscrB\_S-EPIC. 
%%%%%%%%%%%%%%%%%%%%%%%%%%

V.L. F.S. R.F. and N.T. are supported by the Italian Ministry of University and Research (MUR) under the grant PNRR-M4C2-I1.1-PRIN 2022-PE2 Non-perturbative aspects of fundamental interactions, in the Standard Model and beyond F53D23001480006 funded by E.U. - NextGenerationEU.
%%%%%%%%%%%%%%%%%%%%%%%%%%
S.S. is supported by MUR under grant 2022N4W8WR. 
F.S. G.G and S.S. acknowledge MUR for partial support under grant PRIN20172LNEEZ. 
%%%%%%%%%%%%%%%%%%%%%%%%%%
F.S. and G.G.
acknowledge INFN for partial support under GRANT73/CALAT.
%%%%%%%%%%%%%%%%%%%%%%%%%%
F.S. is supported by ICSC – Centro Nazionale di Ricerca in High Performance Computing, Big Data and Quantum Computing, funded by European Union – NextGenerationEU.
\end{acknowledgments}

%%%%%%%%%%%%%%%%%%%%%%%%%%%%%%%%%%%%%%%%%%%%%%%%%%%%%%%%%%%%%%%%%%%%%%%%%%%%%%%%%%%%%%%%%%%%%%%%
\appendix

%%%%%%%%%%%%%%%%%%%%%%%%%%%%%%%%%%%%%%%%%%%%%%%%%%%%%%%%%%%%%%%%%%%%%%%%%%%%%%%%%%%%%%%%%%%%%%%%
\subsection{Appendix: HLT analysis}
\label{sec:appHLT}
\begin{figure*}[!ht]
\includegraphics[scale=0.25]{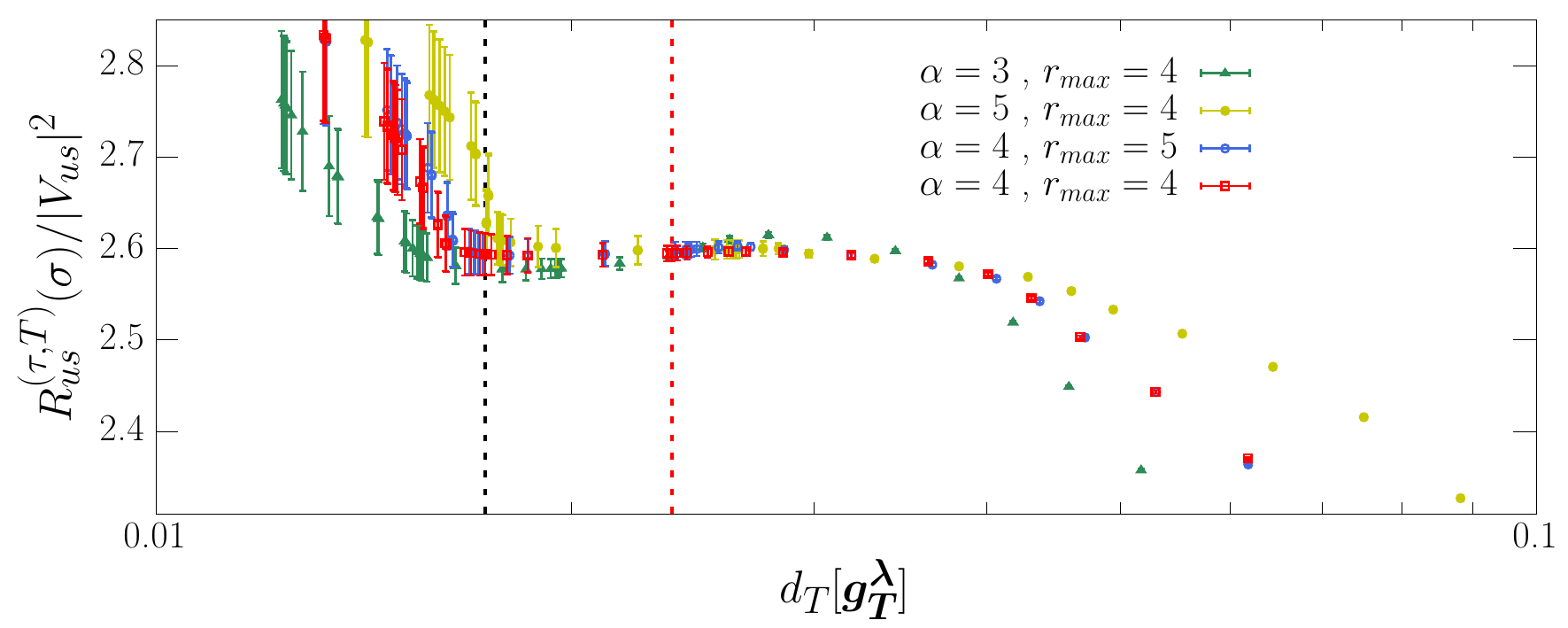}
\includegraphics[scale=0.3]{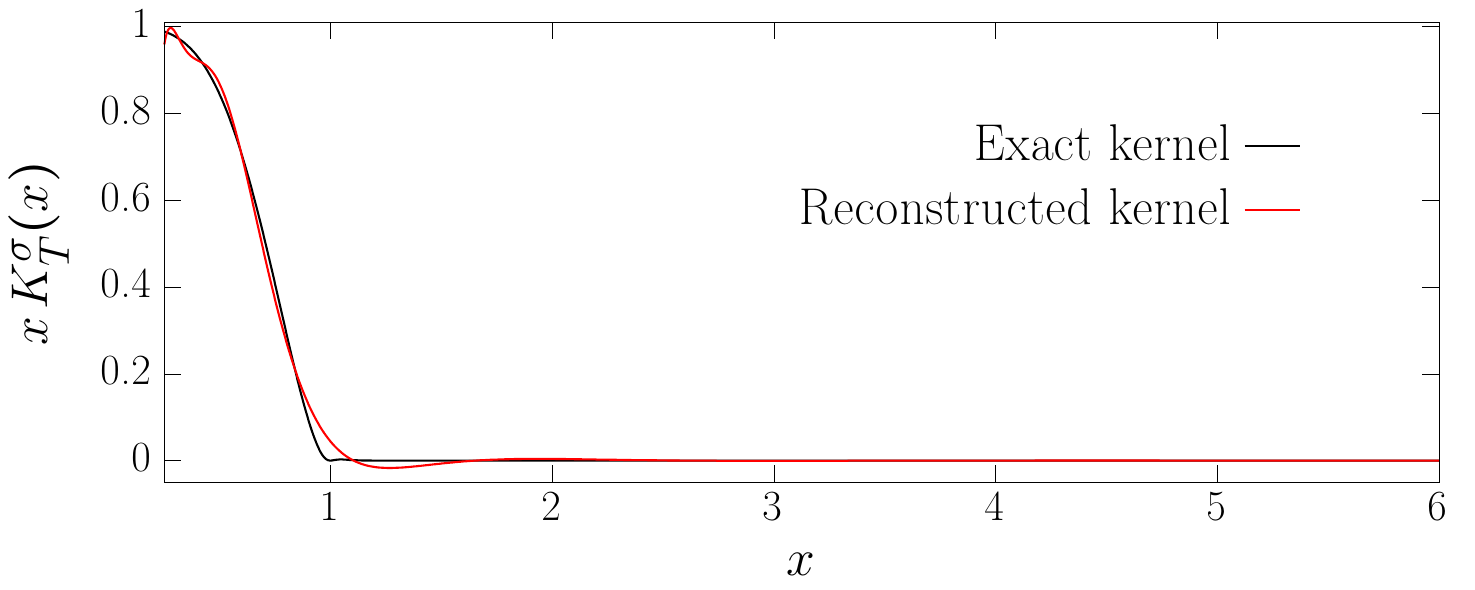} \\
\includegraphics[scale=0.25]{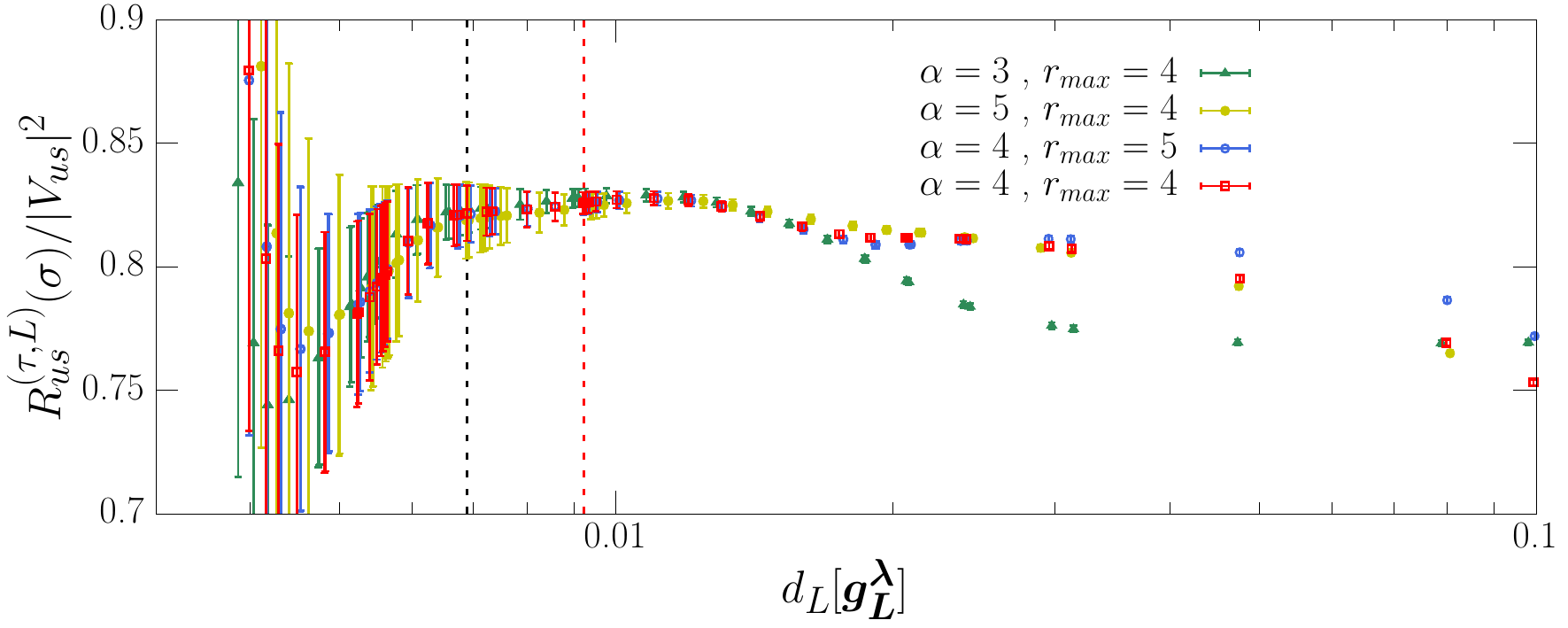}
\includegraphics[scale=0.3]{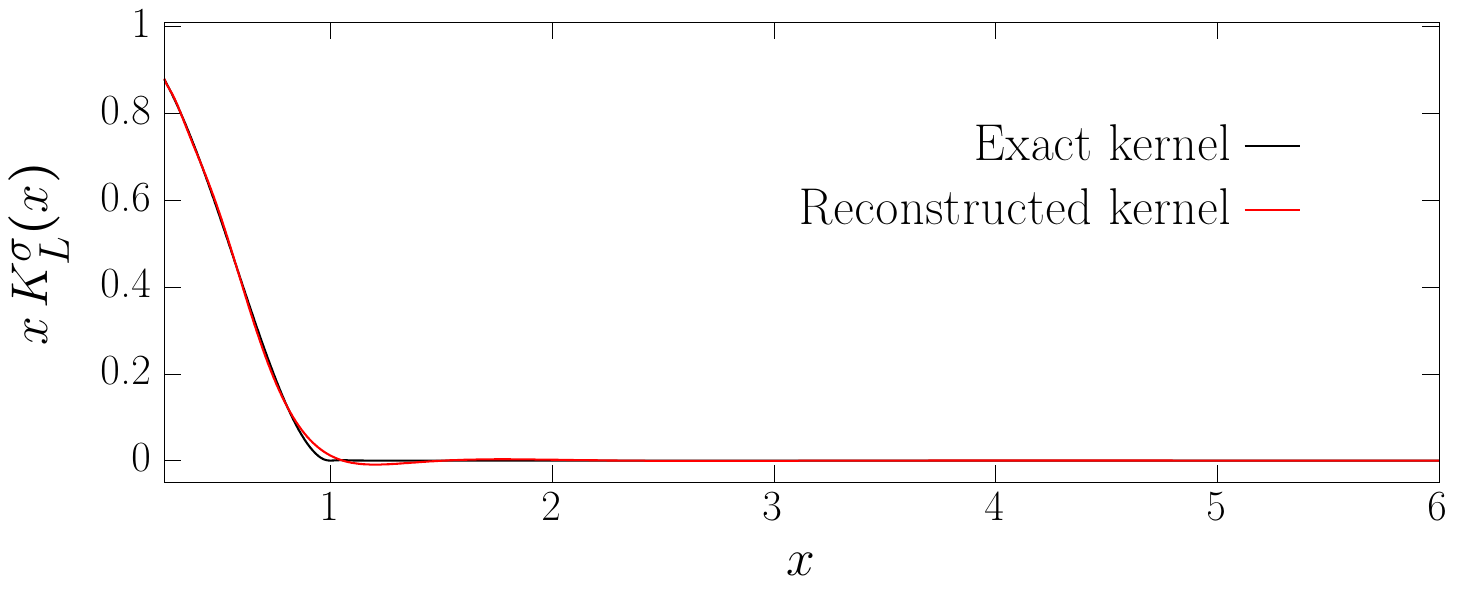}
\caption{\label{fig:stability} \textit{Left:} representative stability-analysis plots for $R_{us}^{(\tau,I)}(\sigma)$. The data are plotted as functions of $d_\mathrm{I}[\vec g_\mathrm{I}^{\lambda}]$ (see Eq.~(\ref{eq:ddef})) and refer to the results obtained on the B64 ($R_{us}^{(\tau,T)}$) and D96 ($R_{us}^{(\tau,L)}$) ensembles for $\sigma= 0.02$, using the ``tm'' and ``OS'' regularization, respectively. In each figure, the points of different colors correspond to different values of the algorithmic parameters $\alpha$ and $r_{max} =aE_{max}$ while the red and black vertical lines to the points $d_\mathrm{I}[\vec g_\mathrm{I}^{\star}]$ and $d_\mathrm{I}[\vec g_\mathrm{I}^{\star\star}]$ that we use to extract the central values and errors of our results. \textit{Right:} comparison between the exact and the reconstructed kernel functions corresponding to the optimal representation obtained in the case $\alpha=r_{max}=4$ (red vertical line in the corresponding left-plot).
}
\end{figure*}
\begin{figure}[h!]
\includegraphics[scale=0.3]{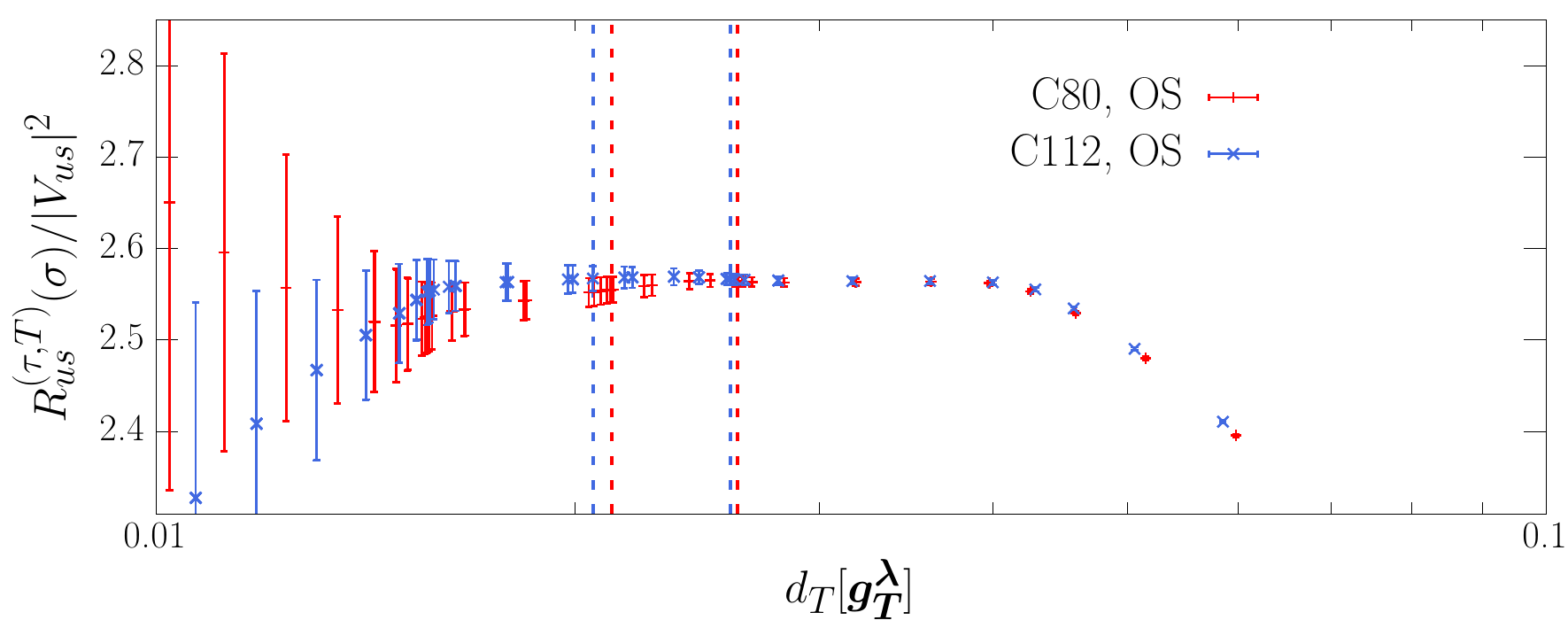}
\includegraphics[scale=0.3]{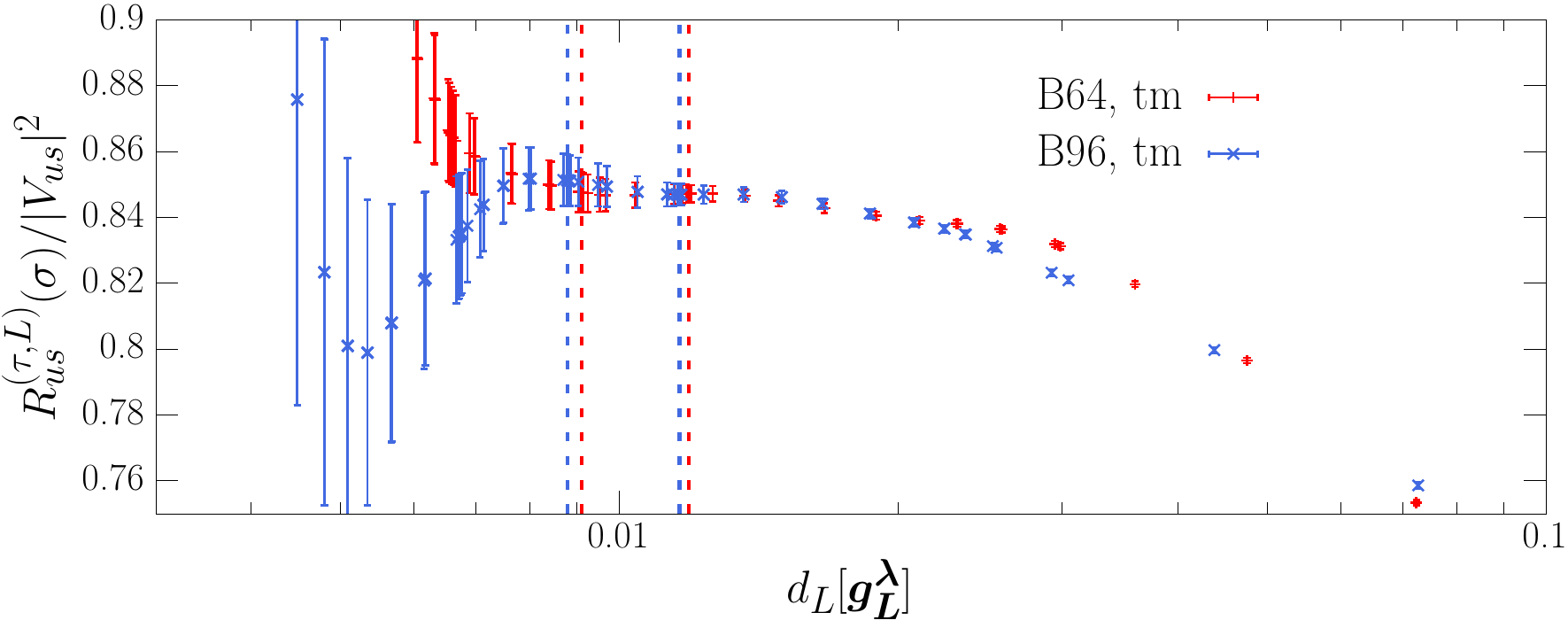}
\caption{\label{fig:FSE} \textit{Top:} comparison between the results obtained for $R_{us}^{(\tau,T)}(\sigma)$ on the C80 and C112 ensembles in the ``OS'' regularization and for $\sigma=0.02$. \textit{Bottom:} comparison between the results obtained for $R_{us}^{\tau,L}(\sigma)$ on the B64 and B96 ensembles in the ``tm'' regularization for $\sigma=0.02$. Data correspond to the reconstruction obtained for $\alpha=r_{max}=4$. FSEs are in almost all cases of similar size as the statistical error. In both panels the vertical lines mark the points $d_\mathrm{I}[\vec g_\mathrm{I}^{\star}] > d_\mathrm{I}[\vec g_\mathrm{I}^{\star\star}]$ for the two ensembles.}
\end{figure}

The HLT coefficients are obtained by considering different definitions of the so-called norm functional,
\begin{flalign}
&
A^\alpha_\mathrm{I}[\vec g_\mathrm{I}]
=
\int_{E_\mathrm{min}}^{E_\mathrm{max}} dE\, e^{\alpha aE} \left\vert
\tilde K_\mathrm{I}^\sigma\left(\frac{E}{m_\tau};\vec g_\mathrm{I}\right) - 
K_\mathrm{I}^\sigma\left(\frac{E}{m_\tau}\right)
\right\vert^2,
\label{eq:Adefs}
\end{flalign}
 measuring the squared distance $\| \tilde K_\mathrm{I}^\sigma- K_\mathrm{I}^\sigma\|^2$  in functional space with different definitions of the norm,
and by balancing the systematic error associated with an imperfect reconstruction of the smearing kernels and the statistical errors on $R^{(\tau,\mathrm{I})}_{us}(\sigma)$. These are proportional to the so-called error functional,
\begin{flalign}
&
B_\mathrm{I}[\vec g_\mathrm{I}]
=
\sum_{n_1,n_2=1}^{N} g_\mathrm{I}(n_1)\,g_\mathrm{I}(n_2)\, \mathrm{Cov}_\mathrm{I}(an_1,an_2)\;,
\label{eq:Bdefs}
\end{flalign}
where $\mathrm{Cov}_\mathrm{I}$ is the covariance matrix of the lattice correlator. At fixed values of the algorithmic parameters $\{N,\alpha,\lambda,E_\mathrm{min},E_\mathrm{max}\}$, the coefficients are obtained by minimizing a linear combination of the norm and error functionals,
\begin{flalign}
\frac{\partial}{\partial g_\mathrm{I}(n)}\left(
A^\alpha_\mathrm{I}[\vec g_\mathrm{I}] +\lambda B_\mathrm{I}[\vec g_\mathrm{I}]
\right)_{\vec g_\mathrm{I}=\vec g_\mathrm{I}^\lambda} =0\;.
\end{flalign}
Given the coefficients $\vec g_\mathrm{I}^\lambda$, the systematic error associated with the approximate reconstruction of the smearing kernel can be quantified by considering
\begin{flalign}
&
d_\mathrm{I}[\vec g_\mathrm{I}^\lambda] = 
\sqrt{\frac{A^0_\mathrm{I}[\vec g_\mathrm{I}^\lambda]}{A^0_\mathrm{I}[\vec 0]}}\;.
\label{eq:ddef}
\end{flalign}
The trade-off parameter $\lambda$ allows to cope with the fact that the statistical errors tend to diverge in the $d_\mathrm{I}[\vec g_\mathrm{I}]\mapsto 0$ and $\sigma\mapsto 0$ limits (see  Refs.~\cite{Bulava:2021fre,Gambino:2022dvu,ExtendedTwistedMassCollaborationETMC:2022sta,Buzzicotti:2023qdv} for extended discussions on this point). Indeed, the quality of the kernel reconstruction improves ($d_\mathrm{I}[\vec g_\mathrm{I}^\lambda]$ decreases) by decreasing $\lambda$, while the statistical errors decrease (at the price of larger values of $d_\mathrm{I}[\vec g_\mathrm{I}^\lambda]$) by increasing $\lambda$. The optimal balance between statistical and systematic errors is obtained by studying the stability of the physical results, i.e.\ of $R^{(\tau,\mathrm{I})}_{us}(\sigma)$ in this case, w.r.t. variations of the algorithmic parameters.  

\paragraph{Stability analyses.}
By relying on the fact that $\rho_\mathrm{L,T}(E^2)=0$ for $E< m_{K}$, we set $E_{\rm min} = 0.9\, m_{K}$. The size of the exponential basis, $N$, has always been fixed by the condition that the uncertainty of $C_{\rm I}(na)$ for $n\leq N$ must be smaller than $10\%$. The algorithmic parameter $E_{\rm max}$ has been set to $E_{\rm max} = r_{\rm max}/a$ and different choices of $r_{\rm max}$ have been employed.  

In FIG.~\ref{fig:stability} we show representative examples of our stability analyses. The robustness of this analysis procedure has been quantitatively assessed in Ref.~\cite{Bulava:2021fre}, where it has been introduced, and in the works~\cite{Gambino:2022dvu,ExtendedTwistedMassCollaborationETMC:2022sta,Barone:2023tbl,Bonanno:2023ljc,Frezzotti:2023nun,Bonanno:2023thi,Evangelista:2023fmt} where it has been subsequently applied. 

On each ensemble and for each regularization the uncertainty on $R_{us}^{(\tau,I)}(\sigma)$ is estimated by varying the parameters of the HLT algorithm and by checking that the results are stable within the statistical errors. The stability plots can be read from right to left (and can be understood in analogy with the more familiar effective-mass plots from which the masses of stable hadrons are usually extracted). In the right-most regions, where $d_\mathrm{I}[\vec g_\mathrm{I}^\lambda]$ is large, the results strongly depend on the choice of the algorithmic parameters (in the analogy these are the  small-time regions in which effective masses are dominated by excited states). The left-most regions, where $d_\mathrm{I}[\vec g_\mathrm{I}^\lambda]$ is very small, correspond to excellent reconstructions of the smearing kernels. In these regions (in the analogy these are the large-time regions in which the signal on the effective masses of nucleons is usually lost) the coefficients $\vec g_\mathrm{I}^\lambda$ tend to become huge in magnitude and oscillating in sign. Neither the central values nor the errors of the results can be trusted when this happens. Indeed, any tiny numerical error on the lattice correlators, even rounding, excludes the possibility of getting trustworthy results for the sums corresponding to Eq.~(\ref{eq:sumresult}) in these cases. A reliable prediction can be obtained when the stability plots show a plateaux in the middle-region (the effective-mass plateaux from which the hadron mass is extracted). By fitting the results in the plateaux region, when it exists, one can certainly get a reliable estimate of the errors. As done in the already referenced  works~\cite{Bulava:2021fre,Gambino:2022dvu,ExtendedTwistedMassCollaborationETMC:2022sta,Barone:2023tbl,Bonanno:2023ljc,Frezzotti:2023nun,Bonanno:2023thi,Evangelista:2023fmt}, we decided here to follow a slightly different procedure that, in fact, provides more conservative estimates of the errors. The central values and the statistical errors of our results are quoted by selecting a first reference point in the stability region, corresponding to $\vec g_\mathrm{I}^\lambda=\vec g_\mathrm{I}^{\star}$ (red vertical lines in FIG.~\ref{fig:stability}). The choice of $\vec g_\mathrm{I}^{\star}$ is obviously not unique. Any point inside the plateaux-region can be selected and the smallest statistical error would be obtained by taking the right-most one. In this work we have chosen the $\vec g_\mathrm{I}^{\star}$ points of the different stability analyses in the middle of the plateaux regions. The systematic errors associated with the necessarily imperfect reconstruction of the kernels have been estimated by selecting a second reference point on the left of $\vec g_\mathrm{I}^{\star}$, that we call $\vec g_\mathrm{I}^{\star\star}$ (black vertical lines in FIG.~\ref{fig:stability}) and that corresponds to the condition
\begin{flalign}
\frac{A^\alpha_\mathrm{I}[\vec g_\mathrm{I}^{\star\star}]}{B_\mathrm{I}[\vec g_\mathrm{I}^{\star\star}]}
=\frac{1}{10}
\frac{A^\alpha_\mathrm{I}[\vec g_\mathrm{I}^{\star}]}{B_\mathrm{I}[\vec g_\mathrm{I}^{\star}]}\;.
\end{flalign}
The ratio between the accuracy of the kernel reconstruction and the statistical errors is 10 times smaller at $\vec g_\mathrm{I}^{\star\star}$ w.r.t. $\vec g_\mathrm{I}^{\star}$. The systematic errors have then been quantified by considering the differences
\begin{flalign}
dR
=
R^{(\tau,\mathrm{I})}_{us}(\sigma;\vec g_\mathrm{I}^{\star}) 
-
R^{(\tau,\mathrm{I})}_{us}(\sigma;\vec g_\mathrm{I}^{\star\star})
\end{flalign}
and their statistical errors ($\sigma_{dR}$) and by weighting these differences by the probability that they are not due to statistical fluctuations, i.e. 
\begin{flalign}
\label{eq:syst_HLT}
\Delta^\mathrm{HLT}_\mathrm{I}(\sigma)
=
\left\vert dR \right\vert
\mathrm{erf}\left(\frac{dR}{\sqrt{2} \sigma_{dR}}\right).
\end{flalign}
We performed 672 stability analyses (one for each ensemble, for each lattice regularization, for each value of $\sigma$ and for each considered value of $\alpha$ and $r_\mathrm{max}$) and found statistical errors typically at the $0.3-0.5$\% level of accuracy and systematic errors larger than three times the statistical errors in only 2.8\% of the cases. The results corresponding to different values of $\alpha$ and $r_\mathrm{max}$ have been used to cross check the reliability of our estimates of the HLT systematic errors after having performed, separately, the continuum and $\sigma\mapsto 0$ extrapolations, see Figure~\ref{fig:sigma_extr} in the main text.

\paragraph{Finite-size effects.}
We carried out a data-driven estimate of the finite-size effects (FSEs), which are quantified by the spread between the results obtained on the C80 ($L\sim 5.5~{\rm fm}$) and on the C112 ($L\sim 7.6~{\rm fm}$) ensembles, weighted by the probability that this spread is not due to statistical fluctuations and maximized over the ``tm'' and ``OS'' regularizations
(see Eqs.~(43) and~(44) of Ref.~\cite{Evangelista:2023fmt}). We then also checked that these estimates are compatible with the corresponding ones coming from the coarser ensembles B64 and B96 and included the B96 ensemble (not corrected for FSEs) as an extra point in our continuum extrapolations. In FIG.~\ref{fig:FSE} we give examples of such comparison. We have found that FSEs are generally small and of similar size as our statistical accuracy (larger than two times the statistical errors in about 1\% of the cases).

%%%%%%%%%%%%%%%%%%%%%%%%%%%%%%%%%%%%%%%%%%%%%%%%%%%%%%%%%%%%%%%%%%%%%%%%%%%%%%%%%%%%%%%%%%%%%%%%
\bibliography{tauvus}% Produces the bibliography via BibTeX.
%%%%%%%%%%%%%%%%%%%%%%%%%%%%%%%%%%%%%%%%%%%%%%%%%%%%%%%%%%%%%%%%%%%%%%%%%%%%%%%%%%%%%%%%%%%%%%%%

\end{document}